\documentclass[a4paper]{jpconf}
\usepackage{hyperref, graphicx, amsmath, amsfonts, epsfig, amssymb}
\def\sg{SGRs/AXPs}
\def\bp{B_{\rm p}}
\begin{document}
\title{Ultra-high energy cosmic rays from white dwarf pulsars and the Hillas criterion}

\author{Ronaldo V. Lobato$^{1,3,4}$, Jaziel G. Coelho$^{2}$ and M. Malheiro$^{1}$}
\address{$^{1}$ Departamento de F\'isica, Instituto Tecnol\'ogico de Aeron\'autica, ITA - DCTA, Vila das Ac\'acias, S\~ao Jos\'e dos Campos, 12228-900 SP, Brazil}
\address{$^{2}$ Divis\~ao de Astrof\'isica, Instituto Nacional de Pesquisas Espaciais - DAS/INPE/MCTI,\\
  S\~ao Jos\'e dos Campos, 12227-010, S\~ao Paulo, Brazil}
\address{$^{3}$ Dipartimento di Fisica, Sapienza Università di Roma, P.le Aldo Moro 5, I-00185 Rome, Italy}
\address{$^{4}$ ICRANet, P.zza della Repubblica 10, I-65122 Pescara, Italy}

\ead{ronaldo.lobato@icranet.org}

\begin{abstract}
 The origins of ultra-high-energy cosmic rays ($E\gtrsim10^{19}$ eV) are a mystery and  still under debate in astroparticle physics. In recent years some efforts were made to understand their nature. In this contribution we consider the possibility of Some Soft Gamma Repeaters (SGRs) and Anomalous X-ray Pulsars (AXPs) beeing white dwarf pulsars, and show that these sources can achieve large electromagnetic potentials on their surface that accelerate particle almost at the speed of light, with energies $E \sim 10^{20-21}$ eV. The sources SGRs/AXPs considered as highly magnetized white dwarfs are well described in the Hillas diagram, lying close to the AR Sorpii and AE Aquarii which are understood as white dwarf pulsars.
\end{abstract}

\section{Introduction}

In our previous work we have raised some issues concerning the origin of ultra-high-cosmic rays (UHECR) \cite{Lobato2016b}. We pointed out that some soft Gamma Repeaters (SGRs) and Anomalous X-ray pulsars (AXPs) could be considered white dwarf pulsars. The idea of SGRs/AXPs as possible sources of the highest energy cosmic rays was investigated by Arons \cite{Arons2003}, that have considered these sources as neutron stars with ultrastrong magnetic field ($B\sim10^{15}$\ G) on their surface, the so-called magnetars. Rotation and strong magnetic fields can induce large potential differences \cite{Aharonian1995, Hooper2009} on the star's surface, accelerating particles with high energy and fulfilling the magnetosphere with pairs of electron and positrons, which turns into a cascade of photons. More recently Kashiyama et al. \cite{Kashiyama2011} have done an important work in this perspective, suggesting that the excess of cosmic ray electron and positrons observed in PAMELA is produced in magnetized White Dwarfs (WDs), by the same mechanism as the one of neutron star pulsars.

In fact, highly magnetized white dwarfs are able to have a large electric field on their surface. Recently, a great number of massive and magnetic white dwarfs ($10^{6}$-$10^{9}$\ G) have been observed \cite{Kepler2013}, and in 2016 a white dwarf pulsar known as AR Scorpii has been discovered with some minutes of period, showing radiation in a broad range of frequencies \cite{Marsh2016, Buckley2017}. Thus, these fast and magnetized WDs can be in principle a potential source for ultra-high-energy-cosmic rays, and in this contribution we will investigate this possibility.

\section{The sources}
SGRs/AXPs are typically identified with very slow rotating pulsars $P\sim(2-12)$ s, with intense magnetic fields $\sim10^{15}$ G \cite{Duncan1992a}. Their spin-down rates $\dot{P}\sim(10^{-13}-10^{-10})$ s/s are larger than the ones of normal pulsars $\dot{P}\sim(10^{-15}-10^{-14})$ s/s \cite{Turolla2015}. The magnetar's X-ray luminosity $L_X$ is explained by the decay of their enormous magnetic field. These sources do not can be understood as rotation powered neutron stars, because their rotational energy is much smaller than X-ray luminosity. Currently there are 23 SGRs/AXPs\footnote{\url{http://www.physics.mcgill.ca/~pulsar/magnetar/main.html}} that are classified as magnetars and 6 until now only candidates \cite{Olausen2014}.

\subsection{SGRs/AXPs as white dwarf pulsars}

It has been proposed by Malheiro, Rueda and Ruffini an alternative model considering SGRs/AXPs as white dwarf pulsars \cite{Malheiro2012}. As discussed in \cite{Usov1988}, the process of release energy for dipole radiation in a white dwarf can be explained in terms of a canonical spin-powered pulsars model, since in certain aspects they are similar (see e.g., \cite{Coelho2013a, Malheiro2013, Coelho2014c}).\\
For example, if we consider a star with $M=1.4{M_{\odot}}$ and $R=10^{6}\ \rm{cm}$, the magnetic field at poles is given by,
\begin{eqnarray}
B_{\rm p}^{\rm NS}=3.2\times 10^{19}(P\dot{P})^{1/2}\rm{G}.
\end{eqnarray}
In the case of a white dwarf with $M=1.4M_{\odot}$ and $R=3\times 10^8\ \rm{cm}$, there is a new scale for the magnetic field at poles,
\begin{eqnarray}\label{bwhite}
B_{\rm p}^{\rm WD}=4.21\times 10^{14}(P\dot{P})^{1/2}\rm{G}.
\end{eqnarray}
Following the last consideration there are new values for the mass density, moment of inertia, dipole moment and rotation energy \cite{Coelho2014c,Coelho2013b}. This description is supported by the observational growth of fast, massive and highly magnetized WDs \cite{Kepler2013,Boshkayev2013a, Boshkayev2014,Das2013}.

\subsection{White dwarf pulsars}

This description was raised by Ostriker \cite{Ostriker1968b, Gunn1969}. Later, the seminal works of Usov \cite{Usov1988a, Usov1993} have discussed the generation of gamma rays by magnetic white dwarfs, showing that WD can produce pairs of electron-positron $e^{\pm}$. Zhang and Gil \cite{Zhang2005} interpreted the transient radio source, GCRT J1745-3009, as a white dwarf pulsar with a period of 77.13 minutes. In this situation the production of $e^{\pm}$ due curvature radiation is below of the pair-production threshold, still if the star is assumed with a field $B_{\rm p}^{\rm WD}=10^{9}$\ G. We highlight the recent discovery, AR Scorpii (AR Sco's), as a pulsating white dwarf in a binary system confirming the hypothesis of white dwarf pulsar \cite{Marsh2016}. AR Sco's has a range of mass $0.81M_{\odot}<M1<1.29M_{\odot}$ and pulses with a period of 1.97 minutes. These pulses reflect the spin of a WD, slowing down on a $10^{7}$ $yr$ timescale. The AR Sco's broadband spectrum is characteristic of synchrotron radiation, requiring relativistic electrons, possibly originate from near the white dwarf and accelerated to almost the speed of light. These particles produce radiation from X-ray to radio wavelengths, typically of neutron star pulsars. This rapidly rotating magnetized WD would simulate the neutron star pulsars as pointed out by Geng et al. \cite{Geng2016}. Another specific example is AE Aquarii, the first white dwarf pulsar, with a short period $P=33$ s and spinning down at a rate $\dot{P}=5.64\times10^{-14}$ s/s. The rapid braking of the white dwarf and the nature of hard X-ray pulses detected with SUZAKU space telescope \cite{Terada2008} can be explained in terms of spin-powered pulsar mechanism \cite{Ikhsanov1998b}. Although AE Aquarii is a binary system with orbital period $∼9.88$ hr, pretty likely the matters' accretion is inhibited by the fast rotation of the white dwarf.

\section{General constrains from geometry and radiation to UHECR}

There are some constrains that UHECR particles should satisfy, as shown by \cite{Ptitsyna2010}:
\begin{itemize}
\item geometry - the accelerated particle should need be kept inside the source while being accelerated;
\item power - the source should posses the required amount of energy to give it to accelerated particles;
\item radiation losses - the energy lost by a particle for radiation in the accelerating field should not exceed the energy gain;
\item interaction losses - the energy lost by a particle in interactions with other particles should not exceed the energy gain;
\item emissivity - the total number (density) and power of sources should be able to provide the observed UHECR flux;
\item accompanying radiation of photons, neutrinos and low-energy cosmic rays should not exceed the observed fluxes, both for a given source and for the diffuse background.
\end{itemize}

\section{Hillas criterion}
If a particle escapes from the region where it was being accelerated, it will not be unable to gain more energy. The one in question establish a limit on maximum energy $E_{\textrm{max}}$ acquired by a particle passing in a medium with magnetic field $B$,
\begin{equation}
  E_{\textrm{max}}=ZqBR_s,
\end{equation}
where $q$ is the electric charge of the particle, $B$ is the magnetic field, $R_s$ is the size of the accelerator and $Z$ the atomic number of the particle (for the case of iron, $Z=26$). This equation considers the Larmour radius of the particle, $R_{\textrm{L}}=E_{\textrm{max}}/(ZqB)\leqslant R_s$. 
This is a general geometrical criterion known as the {\it Hillas criterion} for all types of cosmic ray sources \cite{Hillas1984}.
Neglecting energy losses, i.e., the accelerator is $100\%$ efficient, we see that only the parameters $R_s$ and $B$ can describe the source, showing a relationship between the sources' magnetic field strength and its size.

\section{Particle acceleration}
As we discussed above, there are a great number of magnetic white dwarfs and according Usov \cite{Usov1993} if the surface temperature is less than $10^{4}$ K (see \cite{Caceres2017}, for the thermal emission in fast rotating, highly magnetized white dwarfs), a scale height of its atmosphere is essentially smaller than the radius of white dwarf $R$ and the WD will have a strong electric field $E$ in its magnetosphere,
\begin{equation}
E_{\parallel}=\vec{E}\cdot\vec{B}/|\vec{B}|\neq0.
\end{equation}
This parallel electric field determines a charge distribution, known as Goldreich-Julian charge density~\cite{Goldreich1969},
\begin{equation}\label{gold}
  \rho_{\rm c}=\frac{1}{4\pi}\vec{\nabla}\cdot\vec{E}=-\frac{1}{2\pi c}\vec{\Omega}\cdot\vec{B}.
\end{equation}
Particles in that charge distribution co-rotate with the star until the distance where the linear velocity reaches the speed of light. This region is commonly known as light cylinder, with a radius $R_{l}=c/\Omega$. Near the surface of the star, the intense electric field tear away particles. Ultra relativistic particles flowing out, move along the open magnetic field lines producing curvature radiation,
\begin{equation}
  E_{\gamma}=\frac{3}{2}\frac{\hbar\gamma^3c}{r_{c}},
\end{equation}
where
\begin{equation}
  r_{c}\sim(Rc/\Omega)^{1/2}
\end{equation}
is the curvature radius of the field lines. The perpendicular energy of the particle is rapidly dissipated by synchrotron radiation.
This curvature radiation interact with the magnetic field and produce secondary $e^{\pm}$ pairs, $\gamma+B\rightarrow e^{-}+e^{+}$, leading to a pair creation avalanche. The value of the Lorentz $\gamma$ factor of the $e^{\pm}$ pairs produced, is given by the potential difference in the polar cap region,
\begin{equation}\label{gamma}
  \gamma=\frac{e\Delta{V}}{mc^2}.
\end{equation}
Neutron star pulsars are able to produce electrons with Lorentz factors $\gamma\gtrsim10^{7}$.
There is a gap $h$ above the polar region, and the potential difference becomes \cite{Ruderman1975}
\begin{equation}\label{delta}
  \Delta{V}=\frac{\bp\Omega h^2}{2c},
\end{equation}
where $h$ is given by
\begin{equation}
  h\approx\left(\frac{R^3\Omega}{c}\right)^{1/2}.
\end{equation}

\section{Discussions}
Using the dipole formula given in Ref. \cite{Ferrari1969} to calculate the magnetic field, we have shown that the magnetic fields for SGRs/AXPs as highly magnetized white dwarf are at the order of \cite{Coelho2013b,Coelho2012a,Lobato2016, Lobato2015, Lobato2015a},
\begin{equation}
  \bp^{\rm WD}\sim10^9 {\rm G}.
  \end{equation}
This allows us to estimate the charge density in the magnetosphere, i.e., the Goldreich-Julian density given by equation~\ref{gold},
\begin{equation}
  \rho_{c}^{\rm WD}\sim10^{-2}{\rm cm^{-3}},
\end{equation}
this value is $10^{5}$ smaller than the ones when \sg\ are considered as neutron stars, which have a magnetic field $B\sim10^{14-15}$~G. Thus, in the magnetosphere of \sg\ being white dwarfs, there are less charged particles than the neutron stars ones.

From equation~\ref{delta}, we can calculate the maximum potential difference achieved on the surface of \sg\ as white dwarfs, using $\Omega\sim1$~Hz, $R\sim3\times10^3$~km,
\begin{equation}
  \Delta{V}\sim10^{16}{\rm V}.
\end{equation}
This difference is achieved in a length of $h\sim10^3$ cm, and allows the particles accelerated to reach ultra relativistic energies, larger than that of neutron star pulsars. Therefore, white dwarfs can accelerate electrons with Lorentz factors at least $\gamma\sim10^{10}$, one thousand times larger than those accelerated by neutron star pulsars, being capable to produce curvature photons with an energy up to $\sim10^{21}$ eV.

\begin{figure}[h!]\label{hillas}
\begin{center}
  \includegraphics[width=0.95\textwidth, height=0.55\textheight]{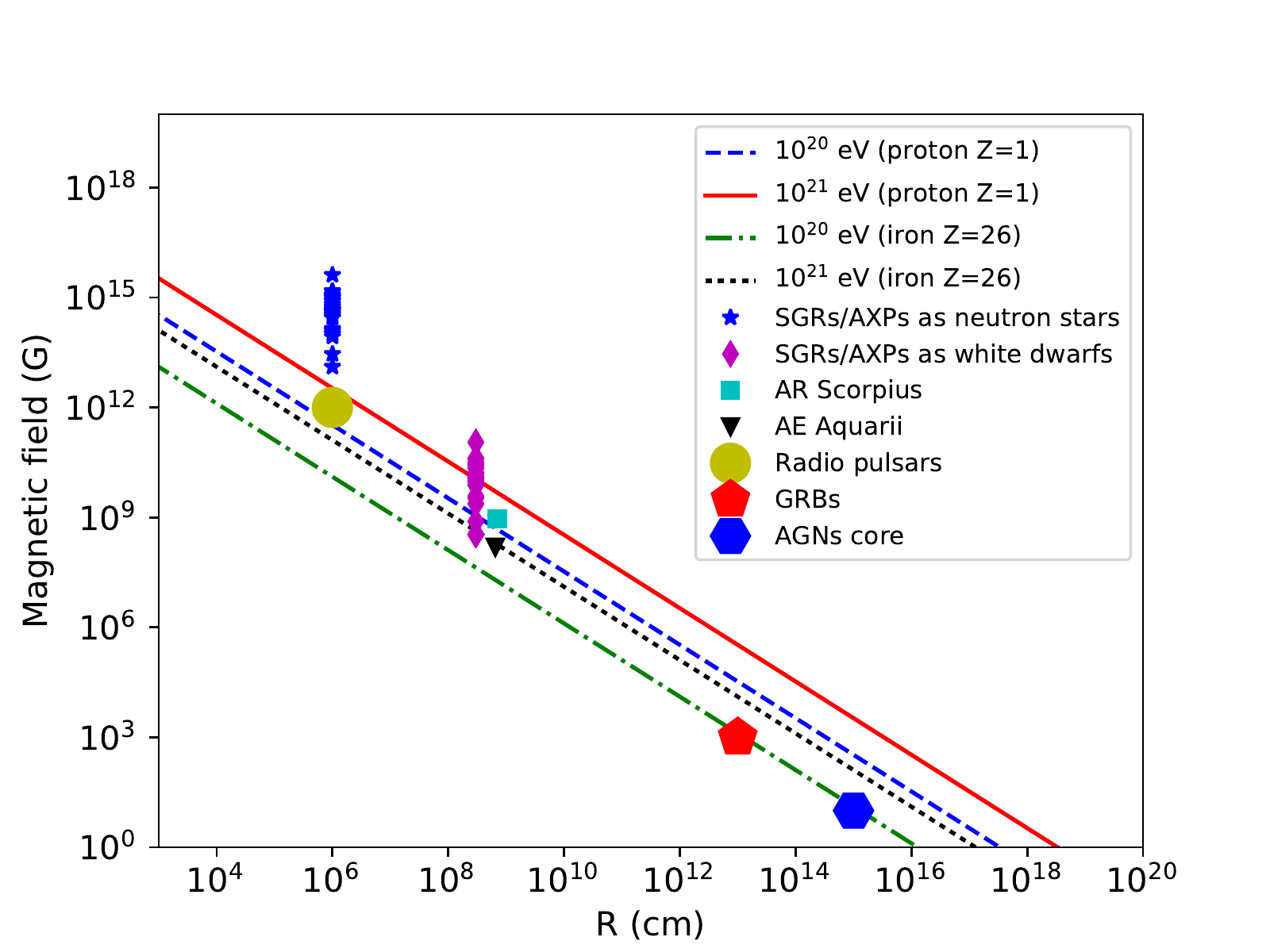}
  \caption{Hillas plot showing the magnetic field strength versus size of SGRs/AXPs as NSs (magnetic field $\sim 10^{13-15}$ G and radius of $10^{6}$ cm) and WDs (magnetic field $\sim 10^{8-10}$ G and radius of $3\times 10^{8}$ cm). The blue-stars and magenta-diamonds describe the SGRs/AXPs as neutron stars and white dwarfs, respectively. Here, we are considering these cosmic ray sources for a maximum energy of $E_{\textrm{max}}=10^{20-21}$ eV for protons (blue and orange lines) and $E_{\textrm{max}}=10^{20-21}$ eV for iron (green and black lines). The cyan-square and black-triangles represent AR Scorpius with magnetic field of $\approx9\times10^{8}$ G and radius of $7\times10^{8}$ cm, and AE Aquarii with magnetic field of $\approx1.5\times10^{8}$ G and radius of $6.5\times10^{8}$ cm, respectively. As we see these two sources are able to accelerate ultra-high energy cosmic rays. It's worth  mentioning other possible sources of cosmic rays pointed in the literature: the yellow-circle, red-pentagon and blue-hexagon are the known radio pulsars, gamma rays-bursts (GRBs) and active Galaxy nuclei (AGNs), respectively.}
\end{center}
\end{figure}

Figure 1 show the Hillas plot for a maximum energy of $E_{\textrm{max}}=10^{20-21}$ eV for protons and $E_{\textrm{max}}=10^{20-21}$ eV for iron. The sources above the green and black lines are able to accelerate atoms of iron up to $10^{20}$ eV and $10^{21}$ eV, respectively. Similarly, the sources above the blue and orange lines are able to accelerate protons up to $10^{20}$ eV and $10^{21}$ eV, respectively. This figure shows that SGRs/AXPs described as WDs and the two known white dwarf pulsars are all on the line of the Hillas plot obtained by Hillas criterion, and consistent with all the others cosmic ray sources known in the universe. Moreover, SGRs/AXPs as magnetars (neutron star pulsars) are out and much above this line, which is not the case for ordinary neutron star pulsars as we also show in Figure 1. Thus, it is quite important to obtain precision measurements of the radius and surface magnetic field of SGRs/AXPs with the new telescopes.

The interaction losses are minimum in the stars' magnetosphere, considering the photon is emitted in a cone with an angular aperture $1/\gamma$ and there is a low charged particle density. Among all SGRs/AXPs, CXOU J010043.1 is the most distant source, located $\sim 62.4$ kpc, thus all sources are within the GZK limit, which is a theoretical upper limit on the energy of cosmic rays coming from outside of our Galaxy ($\approx10$ Mpc for protons with energy of $10^{19}$ eV) \cite{Zatsepin1966,Greisen1966}. Particles accelerated in these sources could be a fraction of ultra-high cosmic ray observed in our planet. All these findings support the plausibility of ultra-high energy cosmic rays from SGRs/AXPs (fast and magnetic massive white dwarfs), in line with important and recent astronomical observations of white dwarf pulsars. We encourage future observational campaigns to verify the radius and the magnetic field of these sources.

\subsection{Acknowledgments}

RVL would like to thank CNPq (Conselho Nacional de Desenvolvimento Cient\'ifico e Tecnol\'ogico) process 141157/2015-1 and CAPES (Coordena\c c\~ao de Aperfei\c coamento de Pessoal de N\'ivel Superior) CAPES/PDSE/88881.134089/2016-01 for financial support. JGC acknowledge the support of FAPESP through the project 2013/15088-0. MM thanks FAPESP thematic project 13/26258-4, CNPq and CAPES.

\section*{References}
\bibliographystyle{iopart-num}
\bibliography{library}
\end{document}